\documentstyle[preprint,tighten,aps,prl]{revtex}
\begin{document}
\draft
\preprint{UTPT-95-14}
\title{Comment on the formation of black holes in nonsymmetric gravity}
\author{ Neil J. Cornish and John W. Moffat}
\address{Department of Physics, University of Toronto \\ Toronto,
Ontario M5S 1A7, Canada}
\maketitle
\begin{abstract}
We critically examine the claim made by Burko and Ori that black holes
are expected to form in nonsymmetric gravity and find their analysis to
be inconclusive. Their conclusion is a result of the approximations they
make, and not a consequence of the true dynamics of the theory. The
approximation they use fails to capture the crucial equivalence principle
violations which enable the full nonsymmetric field equations to detect
and tame would-be horizons. An examination of the dynamics of the full
theory reveals no indication that black holes should form. For these reasons,
one cannot conclude from their analysis that nonsymmetric gravity has black
holes. A definitive answers awaits a comprehensive study of gravitational
collapse, using the full field equations.
\end{abstract}
\narrowtext

\section{Introduction}
Anyone that has looked at alternative gravity theories will have been struck
by just how ubiquitous black holes are. On closer inspection, it becomes clear
why black holes are so hard to get rid of. The experimental success of
Einstein's theory in describing weak gravitational fields requires that all
alternative theories reduce to Einstein's for weak fields. Indeed, for weak
fields this implies that all tenable alternative theories can be recast as
Einstein gravity coupled to some effective matter source constructed from the
additional gravitational degrees of freedom. Now, we know that a black hole
horizon is an entirely regular place so far as local quatities such as
curvatures are concerned. For a large black hole the curvatures can be very
small at the horizon, and we may conclude that the weak field description
of alternative gravity theories should continue to hold. This line of reasoning
suggests that the gravitational collapse of a very massive body will
proceed in much the same way in an alternative theory as it does
in Einstein's theory. For most theories, such as Jordan-Brans-Dicke
scalar-tensor theory, this is exactly what is found.

There is, however, one escape clause in this black hole contract. While it
is true that a free-fall observer sees nothing special at the event horizon,
static observers feel an infinite force. In addition, the redshift between the
horizon and any point outside the horizon diverges. If the alternative theory
violates the equivalence principle or employs non-local notions, then the
horizon can be a very irregular place indeed.

We wish to examine how the preceding considerations apply to nonsymmetric
gravity theory (NGT)\cite{Moff79,john}. In earlier work we showed that the
unique, static, spherically symmetric vacuum solution for NGT did not describe
a black hole\cite{us}.
This result held despite the fact that NGT can be recast as Einstein gravity
coupled to an effective matter source for weak gravitational fields. Recently,
Burko and Ori\cite{lior} considered the lowest-order linearisation of NGT
about an Einstein gravity background and concluded that black holes will form
in
NGT. We show that their conclusion follows as an immediate consequence of the
approximation they used. We stress that the approximation fails to
capture the crucial equivalence principle violations which only occur at
higher orders of approximation. If a horizon is present, the higher order
terms can dominate. By considering gravitational collapse described by the
full NGT field equations we find no reason to expect that black holes will
form.

\section{The linearised theory}

We shall first consider the lowest-order linearised NGT field equations
used to study black hole formation in Ref.\cite{lior}. To first order, the NGT
vacuum field equations linearised about a Einstein gravity (GR) background
read\cite{john}
\begin{eqnarray}
&&R_{\mu\nu}=0 \; , \\
&& \nabla^{\alpha}F_{\mu\nu\alpha}+\mu^2 h_{[\mu\nu]}-4R^{\alpha \; \; \beta}
_{\; \; [\mu\;\;\nu]}h_{[\alpha\beta]}=0 \; ,
\end{eqnarray}
where $F_{\mu\nu\alpha}=h_{\{[\mu\nu],\alpha\}}$ is the field strength formed
from the linearised skew metric $h_{[\mu\nu]}$, and $\mu^2$ is a type of
cosmological constant. At this order, the NGT field equations are identical
to those of a massive, curvature coupled Kalb-Ramond field. The standard
no-hair theorems guarantee that a black hole with $h_{[\mu\nu]}=0$ must form
as the result of the gravitational collapse of matter coupled to such a
Kalb-Ramond field\cite{gary}. The preceding analysis summarises the treatment
presented in Ref.\cite{lior}.

Since violations of the equivalence principle
are restricted to benign curvature couplings at this order, the theory
has been robbed of its ability to see horizons. It is only at higher
orders that interesting violations of the equivalence principle can make
themselves felt in NGT, as we shall explain at the end of this section.

The lowest order equations do correctly predict the breakdown of the
perturbative treatment for a static $h_{[\mu\nu]}$ on a Schwarzschild
background. Of course, this has nothing to do with violations of the
equivalence principle as the same goes for a minimally coupled scalar
field. The breakdown of the linearised analysis requires that the full field
equations be studied. In the case of a scalar field, the pathological
behaviour at the horizon persists in the full field equations, turning the
horizon into a curvature singularity. Thus, before we even begin to study the
collapse of a star coupled to a scalar field we already know what the
outcome must be - a black hole with no scalar hair. In contrast, the full NGT
field equations reveal that a non-zero static $h_{[\mu\nu]}$ is permitted
as curvatures remain small at the Schwarzschild radius and the horizon is
destroyed. Clearly, there is no {\it a priori} reason to exclude the
possibility of a static $h_{[\mu\nu]}$ remaining after a star has collapsed.

We now take a closer look at why we expect higher orders in the skew metric
expansion to introduce important equivalence principle violations. One example
of an effect which only occurs at higher orders involves the volume form.
In GR it is always possible to find a coordinate patch in the neighborhood of
any regular point in terms of which the volume form is identical to that in
Minkowski space. The same is not true in NGT. Importantly, this effect cannot
be seen at first order, so it is an example of an equivalence principle
violating effect missed in the analysis of Ref.\cite{lior}. Another example of
an effect missed at lowest order is due to the local anisotropy of spacetime
caused by the skew field. This local anisotropy alters the propagation
of light\cite{jody}. A related, but more serious effect
is the modification to the propagation of skew perturbations due to non-linear
self-interaction\cite{erwin}. Because of this effect, skew waves will not
follow geodesics of the background geometry in the geometrical optics
approximation. This departure from geodesic motion can lead to divergent
results at the horizon. Such an effect can be important as non-geodesic
trajectories suffer infinite proper accelerations at the event horizon.

We see that the breakdown in the skew perturbation theory can only been
expected {\em when higher-order terms are taken into account and a horizon is
present in the background geometry}. Since the first order analysis fails to
capture vital features of the full theory, we conclude that the first-order
analysis is at best inconclusive, at worst totally misleading. Unfortunately,
the NGT field equations are realted non-polynomially to GR so there are an
infinite number of higher order terms which must be considered. We would have
to prove that divergences do not occur at any order for the first order
analysis to be trusted. Clearly, this is an impossible task so a perturbative
approach using a GR background must be abandoned. The full field equations
must be consulted.

An analogous result has recently been found for string theory in the
presence of horizons\cite{larus}. The intrinsic non-locality of strings
allows them to respond to redshifts. When a horizon is present, the
standard low-energy effective action must be modified to include massive,
extended string modes. The usual description of low energy string gravity
consists of Einstein gravity coupled to a massless dilaton and a massless
Kalb-Ramond field. For weak fields, the higher-order massive modes are
suppressed and can be neglected. The standard dogma states that the horizon for
a large black hole is a weak field region, so we might expect the massless
low energy string theory to continue to be valid when a horizon is present.
In the case of strings, the standard dogma fails because the theory
is non-local. For NGT, the standard dogma fails because the theory incorporates
a special kind of equivalence principle violation. It should be mentioned
however, that the higher order stringy effects considered in Ref.\cite{larus}
are intrinsically non-local, and are not expected to impact on the local
dynamics of gravitational collapse\cite{lp}.

\section{The full theory}
Since the linearised treatment cannot be trusted, we must look at the
full field equations. While we do not claim that the following argument is
rigorous, it does give some idea about what we might expect to find
for gravitational collapse described by the full theory.

The static spherically symmetric metric in NGT can be written as
\begin{equation} \label{met}
g=e^{\nu} dt\otimes dt -\alpha(\nu) d\nu \otimes d\nu
 -r^2 d\theta \otimes d\theta
-r^2 \sin^2\!\theta d\phi \otimes d\phi +f(\nu)\sin\!\theta\,
 d\theta \wedge d\phi \; .
\end{equation}
The radial variable $r$ is a function of $\nu$. For the vacuum Wyman
solution\cite{max} we find that $r(\nu)$ is given implicitly by
\begin{equation}
e^{\nu}(\cosh(a\nu)-\cos(b\nu))^2{r^2 \over 2M^2}=\cosh(a\nu)\cos(b\nu)
-1+s\sinh(a\nu)\sin(b\nu) \; ,
\end{equation}
where
\begin{equation}
a=\sqrt{{\sqrt{1+s^2}+1 \over 2}}\; , \hspace{0.5in}
b=\sqrt{{\sqrt{1+s^2}-1 \over 2}}\; ,
\end{equation}
and $s$ is a dimensionless constant which varies from body to body. At
large $r$ we find
\begin{equation}
r \simeq -{2 M \over \nu} \; .
\end{equation}

It is instructive to consider the massless scalar wave equation
in the metric (\ref{met}). Using the full NGT connection,
we find that the monopole mode of a scalar field obeys
the relation
\begin{equation}
e^{-\nu}\,{\partial^2  \Phi \over \partial t^2}-{ 1 \over \alpha}\,
{\partial^2 \Phi \over \partial \nu^2} = 0 \; .
\end{equation}
When we study this equation on a Wyman background, we see that
small $\Phi$ perturbations remain small everywhere as the background
is everywhere regular. This is true even in the static limit where
the wave equation has the explicit solution
\begin{equation}
\Phi=\Phi_{0}+\Phi_{1}\nu \; .
\end{equation}
Since the maximum value for $\nu$ is roughly $\pi /s$, we can always
choose $\Phi_{0}$ and $\Phi_{1}$ so that $\Phi$ is small everywhere.
The NGT vacuum solution admits scalar hair.
This is in stark contrast to the situation in GR where
\begin{equation}
\Phi=\Phi_{0}+\Phi_{1}\ln\left(1-{2M \over r}\right) \; ,
\end{equation}
and we must set $\Phi_{1}=0$ to obtain a solution regular at $r=2M$.

A similar, but far more complicated, analysis can be made for
skew metric perturbations about the metric $(\ref{met})$. Indeed, these
kind of skew metric perturbations about the static solution have to be
included as NGT does not have an analog of Birkoff's theorem. The
skew metric perturbations lead to a set of coupled hyperbolic
differential equations. This is because a first order skew perturbation
excites first order perturbations in the symmetric metric functions. We
note that this is in contrast to what we found for skew perturbations about a
GR background, where the skew and symmetric sectors remain uncoupled at first
order. Despite these technical complications, the physical picture is the
essentially the same as what we have described for scalar perturbations. Since
the background metric is everywhere regular, small skew perturbations will
remain small.

The same holds for skew perturbations about a star described by NGT. The NGT
metric inside the star is regular, and it matches smoothly onto an exterior
Wyman vacuum solution. Small skew perturbations about the static solution
remain small. This continues to be the case if the star undergoes gravitational
collapse. The endpoint of collapse is likely to be some kind of matter
distribution supported by the repulsive skew fields and matter pressure.

Since skew perturbations on top of the full NGT metric for a collapsing
star are expected to remain small, it is difficult to see how a black hole
might
form. This is because the end-state of spherically symmetric gravitational
collapse in GR, the Schwarzschild metric, differs from the NGT vacuum metric by
a non-perturbative amount. Thus, we require non-perturbative skew fluctuations
to occur if we wish to recover a black hole. As the linearised skew
perturbations
show no sign of diverging, the required non-perturbative skew fluctuations are
ruled out.

\section{Conclusions}
We have pointed out that the lowest order linearisation used in Ref.\cite{lior}
cannot be trusted when the background geometry contains a horizon. For this
reason, we find the analysis in Ref.\cite{lior} to be inconclusive, and the
claim that black holes form in NGT to be premature. Moreover, a parallel
treatment using a NGT background, rather than a GR background, leads to the
opposite conclusion - black holes will not form in NGT. The only way to
really find out whether or not black holes form in NGT is to solve the
collapse problem using the full NGT field equations. Until that is done
properly, the jury is out.

\section*{Acknowledgments}
We thank Lior Burko and Amos Ori for informing us of their work prior
to publication, and for discussing the problem with us.


\begin{references}
\bibitem{Moff79} Moffat J. W., {\em Phys. Rev.} D {\bf 19}, 3554,
1979.
\bibitem{john} Moffat J. W., {\em J. Math. Phys.} to appear, 1995;
{\em Phys. Lett.} {\bf B}, to appear, 1995.
\bibitem{us} Cornish N. J., Moffat J. W., {\em Phys. Lett.}
{\bf B336}, 337, 1994; {\em J. Math. Phys.} {\bf 35}, 6628, 1994.
\bibitem{lior} Burko L. M. and Ori A., {\em Technion Preprint}
TECHNION-PH-95-7,
gr-qc/9506033, 1995.
\bibitem{gary} This is not true when $\mu=0$ and there is no curvature
coupling. Such gauge-invariant Kalb-Ramond fields may give a black
hole some hair, Bowick, M. J., Giddings, S. B., Harvey, J. A.,
Horowitz, G. T., and Strominger, A., {\em Phys. Rev. Lett.} {\bf 61}, 2823,
1988.
\bibitem{jody} Gabriel, M. D., Haugan, M. P., Mann, R. B., and Palmer J. H.,
{\em Phys. Rev. Lett.} {\bf 67}, 2123, 1991; {\em Phys. Rev.} D {\bf 43},
308, 1991; {\em Phys. Rev.} D {\bf 43}, 2465, 1991.
\bibitem{erwin} Hittmair, O., and Schrodinger E., {\em Comm. Dublin Inst. Adv.
Studies}, A{\bf 8}, 1951.
\bibitem{larus} Lowe, D. A., Polchinski, J., Susskind, L., Thorlacius, L.,
and Uglum, J. R., {\em Preprint}, hep-th/9506138, 1995.
\bibitem{lp} Thorlacius, L., {\em Private communication}.
\bibitem{max} Wyman M., {\em Can. J. Math.} {\bf 2}, 427, 1950.
\end{references}
\end{document}